\documentstyle[psfig]{aa}
\begin{document}
\thesaurus {10.05.1, 10.07.1, 10.07.2, 10.11.1, 10.19.3, 11.19.4}

\title{ N-body simulations of globular cluster tides}
\author{Fran\c coise Combes\inst{1}, St\'ephane Leon\inst{1,2} and
Georges Meylan\inst{3}}    

\offprints{F.~Combes \hfill\break(e-mail: bottaro@obspm.fr)}   

\institute{DEMIRM, Observatoire de Paris, 61 Av. de l'Observatoire,     
 F-75014 Paris, France
\and
Institute of Astronomy \& Astrophysics,  Academia Sinica, P.O. Box 1-87, 
Nankang, Taipei, Taiwan 
\and
ESO, Karl-Schwarzschild-Str. 2, D-85748 Garching, Germany}

\date{Received June 7, 1999; accepted October 7, 1999}

\maketitle
\markboth{F. Combes, S. Leon \& G. Meylan:
Globular cluster tides }{}

\begin{abstract}
We present N-body  simulations of globular  clusters, in orbits around
the Galaxy, in order to   study quantitatively and geometrically   the
tidal  effects they   encounter.   The  clusters  are modelised   with
multi-mass  King-Michie   models   (Michie 1963),    including    mass
segregation at initial conditions. The Galaxy is modelled as realistic
as possible, with three components:  bulge, disk and  dark halo.   The
main  finding is  that there exist  two giant  tidal tails around  the
globuler    cluster in  permanence  along  its   orbit,  whatever this
orbit. The length of these tails is of the order  of 5 tidal radii, or
greater. The escaped stars are distributed  radially as a power law in
density, with  a  slope of --4.  The  tails  present substructures, or
clumps, that are   the relics of  the  strongest shocks.    Due to the
compressive disk-shocking, the clusters  display a prolate shape which
major   axis is   precessing around  the    z  axis.  The  tails   are
preferentially formed  by the lowest mass  stars, as expected, so that
the tidal truncation increases  mass segregation. Internal rotation of
the cluster  increases the  mass loss. The   flattening of dark matter
cannot  influence  significantly the  dynamics  of  the clusters.  The
orientation and the strength of the tidal tails  are signatures of the
last    disk crossing, so  that   observed  tidal tails can  constrain
strongly the  cluster orbit and  the galactic model (vertical scale of
the disc).


\end{abstract}

\section{Introduction}

Globular clusters are  fascinating  systems, since contrary to   their
apparent geometrical  simplicity, they are the  sites of  many complex
physical phenomena. The two-body relaxation time  is on average of the
order  of 10$^9$  yr,   shorter  than their  life-time,  so  that  the
relaxation is very efficient especially in the  core, where the memory
of the cluster's initial conditions is expected to  be washed out (see
the review by Spitzer 1987).  The result of this  relaxation is a slow
collapse of   the core, while the less   bound stars in   the envelope
evaporate. Moreover, the  relaxation tends to establish  equipartition
of energy,  and  mass  segregation, so  that  the low-mass  stars  are
preferentially expelled into  the envelope.  Mass  loss due to stellar
evolution is no longer significant for the old clusters in our Galaxy,
but internal dynamical evolution alone  could destroy a large fraction
of the   population    (H\'enon 1961).  In  terms    of  the half-mass
relaxation time t$_{rh}$, the collapse of the core occurs in $\sim$ 15
t$_{rh}$, while total evaporation  occurs in $\sim$ 100  t$_{rh}$. The
core collapse is however unspectacular, involving less than 1\% of the
stars; it can be reversed,  and gravothermal oscillations can  happen,
according to the amount of heating provided  by binaries (Makino 1996,
Kim et al. 1998).

External perturbations can  considerably  accelerate the evolution  of
globular clusters: compressive shocks at the  crossing of the galactic
plane,  tidal  interaction with the  bulge, or  the dark  matter halo.
These external perturbations do much more than a tidal limitation of a
cluster   in a  circular  orbit,  corresponding to  a time-independent
external potential; they are also cause of tidal stripping and heating
of  the cluster (Allen  \&  Richstone 1988).  Paradoxically, they also
accelerate the core collapse (Spitzer \& Chevalier 1973).  Much effort
has been devoted  to quantify  the  dynamical evolution of   clusters,
since it is   crucial to be able to   deduce their initial  number and
distribution and  to  go back to  the Galaxy  formation. This has been
done with   the Fokker-Planck  method, orbit-averaging the  relaxation
effects,  and    estimating    the  tidal   shocks   through   impulse
approximations  (Chernoff et al. 1986,  Aguilar et al.  1988).  It was
already  concluded that the number of  remaining globular clusters now
is a  small  fraction  of  those   formed initially (Aguilar   et  al.
1988). The  majority  of  them were   destroyed  in  the early  Galaxy
evolution, through   violent    dynamical interactions.  Kundi\'c   \&
Ostriker (1995)  and Gnedin  \&   Ostriker (1997) have revised   these
estimates, showing   that the second-order   tidal shocking  ($<\Delta
E^2>$) is even more important than the first order ($<\Delta E>$); the
corresponding shock-induced  relaxation   could dominate the  two-body
relaxation.   It   ensues  that about 75\%   of   all present globular
clusters will   be  destroyed  in the next   Hubble   time,  which  is
compatible   with observational  estimates  (Hut \&  Djorgovski 1992).
Since  already the more  fragile are missing  today, and in particular
there is a depletion of clusters orbiting  within the central 3 kpc of
the galaxy, it is likely that the  initial cluster population was more
than an order of magnitude more numerous than today.

Many uncertainties remain  when trying to   quantify the present  mass
loss of globular clusters.  The effect  of irregularitites in the disk
potential on the clusters evolution have been studied: Giant Molecular
Clouds (Chernoff  et al.   1986),  spiral arms  and bars  (Ostriker et
al. 1989, Long et al. 1992) are found to be only a secondary effect in
the  destruction of the clusters, the  crossing  of a thin plane being
the dominant  factor (and also the bulge  crossing in the inner parts,
Nordquist    et  al. 1999).    Multi-mass   models undergo  more rapid
evolution than single-mass models (Lee  \& Goodman 1995): the rate  of
mass loss  can be  doubled  per half-mass relaxation  time.   Internal
rotation of clusters, that was more important in  the past, might be a
significant factor, too.

Recently  Grillmair  et al. (1995) observed   the  outer parts of 12
Galactic  globular clusters, using  deep  two color  star counts. They
discovered   huge  tidal tails,    consisting  of  stars  escaping the
clusters, that can  help to quantify the mass  loss, and to bring some
constraints on  the cluster orbits.  We have also  carried out  such a
photometric  study on 20  Galactic  clusters (Leon  et al.  1999), and
report about the characteristics  of the  tidal  tails, once  the main
observational biases   are taken    into account (extinction,   galaxy
clusters,  etc...).   In the present   work,  we try  to reproduce the
observations,  in  order  to  better quantify  the  effect of external
perturbations from the Galaxy  on clusters, with mass  segregation and
rotation  included.  In particular we   want to  identify the  fate of
escaped stars, and  relate the tidal tails  morphology  to the cluster
orbits.  In Section 2,  we first describe  the methods  used, together
with  the  models adopted  for the  globular  clusters  and the Galaxy
potential, and describe  the results of  the simulations in Section 3.
Section 4 summarizes the results.

\section{Numerical methods}

\subsection{Overwiew}

One  of  the   most important  problem  in  simulating  the  dynamical
evolution of globular   clusters,  is the  wide  range of  time-scales
involved.  Two kinds   of methods have  been  used:  either an  N-body
integration  with various  algorithms, following the  internal stellar
orbits, which have a dynamical time of the order 1 Myr; but the method
is expensive since the total simulation must be carried out over Gyrs;
also  the two-body  relaxation   might be  over-  or  under-estimated,
according to the  number     of particles used  and   the   softening.
Alternatively, the Fokker-Planck or Monte-Carlo methods are used, with
orbit-averaging and use of diffusion coefficients to take into account
the   two-body relaxation  effects.  But  then external  gravitational
perturbations cannot be    evaluated exactly, and   approximations are
used,   as   adiabatic   invariants,  impulse  approximation,   steady
potentials, etc...

Weinberg (1994a,b,c)   has shown  that the   adiabatic  approximation,
consisting in neglecting the effect of  perturbations slow enough with
respect  to the internal stellar periods,   is generally not valid for
stellar systems, that are heated by slow perturbations. Widely used is
also the impulse approximation, that assumes a very rapid perturbation
(or shock) with  respect to internal  motions; however, this is only a
crude estimate  (Johnston   et al.  1999). Adiabatic  corrections  are
required, and are rather critical, since they can multiply the cluster
destruction time by a factor two (Gnedin et al.  1998, 1999, Gnedin \&
Ostriker 1999).

Oh et al.  (1992a,b, 1995)  developped  an  hybrid method, using   the
Fokker-Planck equations for the cluster center, monitoring the effects
of  two-body relaxation,    and  a  three-body integration    for  the
envelope. They have  been able to  follow  escaped particles  up to 10
initial limiting  radii after  20  orbits around the  galaxy. Recently
Johnston et al.  (1999) used  the  SCF (self-consistent field)  method
developped by   Hernquist \&  Ostriker   (1992), allowing  to simulate
clusters with the actual number of stars (of the order of 10$^6$). The
Poisson noise   is therefore  exactly  natural,  however  the two-body
relaxation    is under-estimated,  except    in  rare   cases  when  a
Fokker-Planck diffusion scheme is added to the N-body code. For a long
time,    Fokker-Planck  calculations  appeared to    yield  much lower
lifetimes  for   clusters   in   the  Galaxy,   compared     to N-body
simulations. Both methods are now  converging (Takahashi \&  Portegies
Zwart 1999).

Our  aim here is to  determine and quantify  the tidal  effects of the
Galaxy on a globular  cluster along one orbit,  on a time-scale of the
order of t$_{rh}$;  we do not focus on  long-term effects, and  do not
follow the cluster  until its destruction.  The two-body relaxation is
only approximated, and we do not take into account dynamical friction,
which occurs on very   long time-scale and  for very  massive clusters
only. We assume the globular cluster old enough so that the effects of
stellar  evolution are negligible.  We  however take  into account the
mass  segregation inside  the cluster,  since it  can affect the tidal
tail behaviour  or the mass  loss, most  of the  small stellar  masses
being    confined  in   the outer   parts.    We will   focus  on  the
characteristics of the tidal tails, their amplitude, and 3D shapes, in
order to  compare with observations,  and therefore to put constraints
on the globular clusters orbits and mass loss.

\subsection{Algorithm}

The N-body  code used is an  FFT algorithm, using  the method of James
(1977)  to avoid the  periodic images.   This method finds  correction
charges  on the 2D boundary surfaces,  which,  once convolved with the
Green  function, cancel out  the    effect of images.   It   increases
considerably the efficiency of the FFT method, especially in 3D, since
it avoids  to multiply by  8 the volume where the  FFT is computed. We
used a 128$^3$ grid with  N=1.5 10$^5$ particles, which required  2.7s
of CPU per time  step on a Cray-C94. The   Green function used is  the
g$_2$ function from James (1977), so that the deviation from Newtonian
law is of the order of $R^{-5}$, at  large distance $R$; the resulting
softening parameter is of the order  of the grid size,  i.e. 1pc.  The
units  used   in the simulations are    pc, km/s,  Myr,  and  G=1 (the
corresponding unit of mass is then 2.32~10$^2$ M$_\odot$).

\subsection{Cluster model}

We  used  several   initial  cluster models,  built  from  Michie-King
multi-mass  distributions. We divide the   particules in 10 mass bins,
distributed  logarithmically.   The stellar  masses  in   old globular
clusters range from 0.12 to 1.2  M$_\odot$ essentially, so we simulate
a mass spectrum over one  decade. Although the mass turn-over is
around 0.8 M$_\odot$ for such old star populations, one should also take 
into account the remnants of massive stars that contribute to replenish 
the high-mass end. We adopt  the Salpeter mass function
for   the spectrum, i.e.   $dN/dm  \propto  m^{-2.35}$.   With such  a
spectrum, which  represents quite  nicely the observations  (Richer et
al. 1991), the total number of stars is of the order of 1.2 10$^6$ for
a cluster  mass of  3 10$^5$ M$_\odot$   (or an  average mass  of 0.24
M$_\odot$).  Since we use a constant  total number of particles of 1.6
10$^5$ for all  our models, whatever  their total mass, there will not
be exact correpondence between  particles and stars, but particles are
statistically representative of  the stars.  Therefore, we  ignore the
possible depletion of  stars at both  ends, to only consider power-law
mass spectra within the two mass limits.

\begin{figure}
\psfig{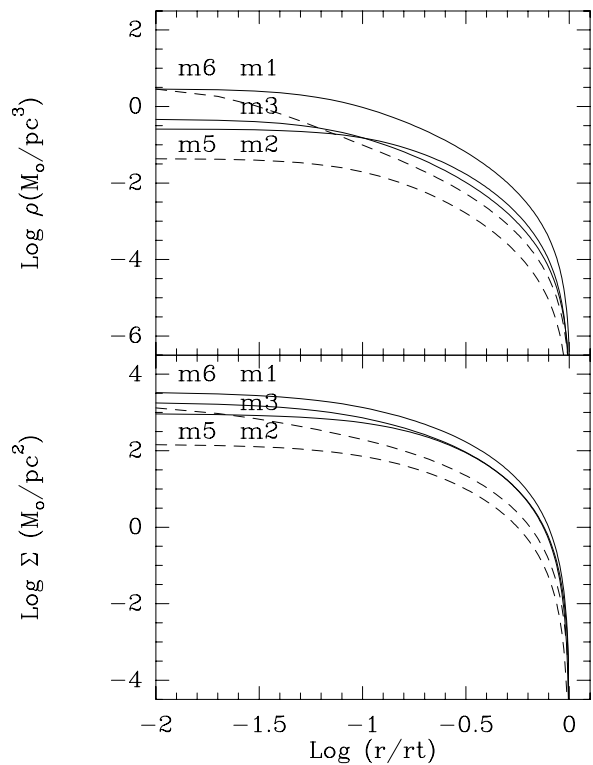}
\caption{ Volumic (upper panel) and surface (lower panel) radial total
density distributions for five of  our  globular cluster models.   The
dotted lines correspond to models m5 and m6 (see Table 1.) }
\label{rho_ini}
\end{figure}

\begin{figure}
\psfig{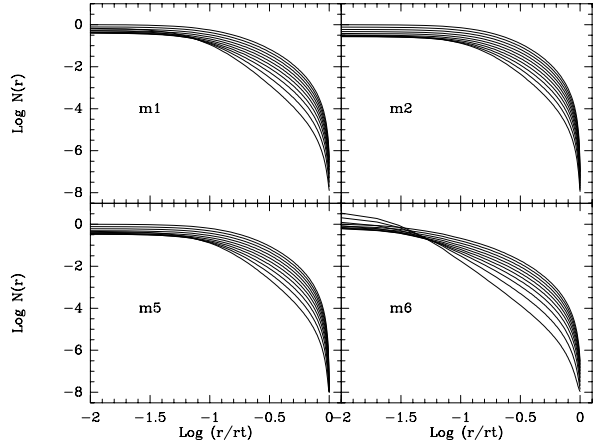}
\caption{ Radial  density law for  the 10  mass bins  for  some of the
globular cluster models. The highest  curve corresponds to low masses,
they are all normalised  to it.  For the  case m6, owing to the higher
concentration, the lower mass stars are less numerous in the center.}
\label{seg_ini}
\end{figure}

To  find the distribution function  for each  mass class, we integrate
the Poisson   equation iteratively, with  the  method described  by da
Costa \& Freeman (1976).   The starting solution  for each mass is the
single-mass distribution function
$$
f(\epsilon) = \rho_0 (2\pi \sigma_0^2)^{-3/2} ( e^{\epsilon/\sigma_0^2} - 1)   
\quad \quad \epsilon > 0
$$
where  $\epsilon  = \Psi -   \frac{1}{2} v^2$ and  $\Psi(r) =- \Phi(r)
+\Phi(r_t)$, $\Phi$  being  the gravitational potential,  $\sigma_0$ the
central velocity dispersion and $r_t$ the tidal radius.  Each model is
determined by  three initial parameters, the  King  core radius r$_0$,
the depth of the potential $W_0$, and  the central velocity dispersion
$\sigma_0$.  The  central density  $\rho_0$  is then derived through the
relation:
$$
r_0^2 = {{9 \sigma_0^2 }\over {4 \pi G \rho_0}}
$$
The velocity dispersions for  each  mass class is  determined  through
equipartition of  energy.  Only a  few iterations  are  required for a
relative accuracy  of 10$^{-3}$, and the  resulting solution gives the
total  mass,  the limiting   radius,  and  the  final  radial  density
distributions   plotted in  Fig.~\ref{rho_ini}.    The degree of  mass
segregation can be estimated from Fig.~\ref{seg_ini}.

\subsection{Rotation}

Rotation is at  present very weak in  globular clusters.  It has  been
measured convincingly only in the brightest cluster, $\omega$ Centauri
(Meylan \& Mayor  1986, Merritt et al.  1997),  where the rotation  is
almost solid-body until about 15\% of the tidal radius, and then falls
off quickly.   It seems that  rotation  is correlated  with flattening
(Meylan \& Mayor 1986, Davoust  \& Prugniel 1990), which is compatible
with  the  result  of  "isotropic oblate  rotator"  found  in $\omega$
Cen. The average observed flattening is  $b/a$ = 0.9, but an important
fraction of the  clusters have axis  ratios smaller than that, between
0.8-0.9.  Since  two-body  relaxation is efficient  in the   core, and
erases all possible primordial anisotropy,  the flattening of globular
clusters is likely due to rotation, contrary to elliptical galaxies.

The influence of rotation  on the dynamical  evolution of clusters has
been investigated by Lagoute  \&  Longaretti (1996) and Longaretti  \&
Lagoute (1997a,b). The  rate of evaporation is increased significantly
(up to a factor of 3 to 4) per relaxation time, although the latter is
somewhat lengthened.  Stellar  escape reduces  the amount of  rotation
and flattening, a result  compatible with the observation of  decrease
of flattening with  age (Frenk \& Fall  1982).  Globular clusters with
shorter relaxation time are  also rounder (Davoust \& Prugniel  1990),
which supports the loss of rotation with relaxation.

Gravitational  shocks   at   disk   crossing  produces  an   apparent
flattening,  mainly parallel to the galactic  plane. It  has long been
argued that the  observed flattening could not  be due to the galactic
tidal field, because  its direction is not  aligned with the  galactic
center (see  Lagoute \&  Longaretti   1996);  but  this  is not    the
expectation  for compressive shocks.   Also, it is possible that tidal
interactions increase the rotation of the  clusters. These effects are
investigated below.

\begin{figure}
\psfig{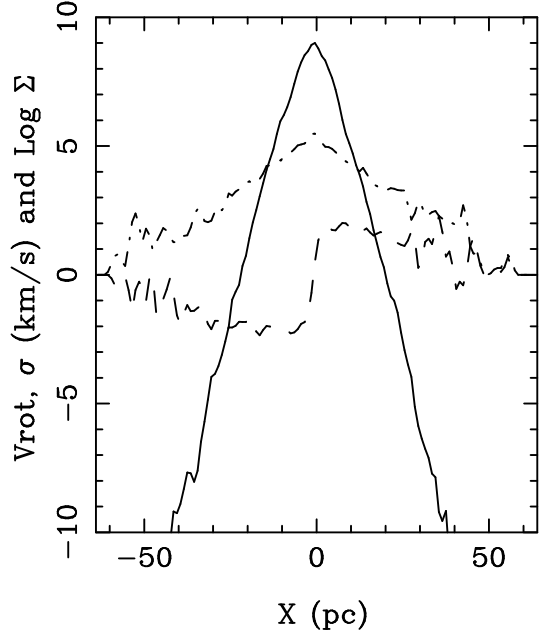}
\caption{ Rotational   profile  of the  cluster model  m2 (dash-line),
velocity dispersion  profile (dash-dot)  and projected density profile
(full line), as  would be observed in  projection perpendicular to the
rotation axis.}
\label{roprof}
\end{figure}

To incorporate rotation in the  initial cluster models, a distribution
function $f(\epsilon,   L_z)$ should be    chosen, corresponding to  a
flattened  density  distribution  $\rho(r,z)$;  however,  only  a  few
studies   have  developped in  this  domain  (cf Wilson 1975, Dejonghe
1986), and no  analytic function has  been found corresponding to  the
likely rotation  curve of clusters. The density  is only a function of
the  even distribution  $f_+ =   0.5 (f(\epsilon,  L_z)  + f(\epsilon,
-L_z))$, since of course the sense  of rotation does not influence the
spatial density, so  that for a given  flattened cluster,  an infinite
distribution  functions could be chosen,  the odd function  $f_- = 0.5
(f(\epsilon, L_z)  -  f(\epsilon, -L_z))$  indicating  the  rotational
velocity.   Since  the rotation  is only a  very weak  effect, we have
selected another scheme to  introduce it. From a non-rotating  cluster
model,  we  select a certain   fraction of the  particles, and reverse
their sign of velocity  ( ${\bf v}$  in $-{\bf v}$) if their projected
angular momentum  $L_z$ is  not  positive.  This  process consists  in
introducing and   $L_z$-odd  part $f_-$  in the  initial distribution.
This  algorithm   allows to control the    amount of  rotation  by the
fraction selected, as a  function  of radius. In  practice,  selecting
particles whatever their radius already   results in a rotation  curve
very compatible  with that  observer for  $\omega$ Cen by  Merritt  et
al. (1997). The cluster is not exactly in equilibrium at start, but is
left to violently   relax (varying potential)  during a  few  crossing
times ($\sim$ 10 Myr), and reaches  quickly a flattened relaxed state,
with flattening   of the order $b/a  \approx  0.9-0.95$. The resulting
rotation profile is displayed in Fig.~\ref{roprof}.

Table  \ref{gct}   displays the principal   parameters  of the cluster
models: the mass,  the  concentration c  = log ($r_t/r_c$),  the core
radius $r_c$ where the surface density is halved, the half-mass radius
$r_h$, the tidal radius $r_t$ which is here the limiting radius of the
King-Michie model initially, the depth of the central potential $W_0 =
\Psi_0/\sigma_0^2$,  the  central number density of stars  n$_0$, the
central  relaxation time t$_{r0}$,  and the  half-mass relaxation time
t$_{rh}$, as defined by Spitzer (1987):
$$
t_{r0} = {{0.065 v_m^3} \over {n_0 m^2 G^2 ln\Lambda} }
$$
and
$$
t_{rh} = 0.138 {{N^{1/2}r_h^{3/2}} \over {m^{1/2}G^{1/2}ln\Lambda}} =
{ {1.7 10^{-4} r_h(pc)^{3/2} N^{1/2}} \over {m(M_\odot)^{1/2}} }  Gyr
$$
where  $N$ is the true  star  number, $m$ the  mean  stellar mass, and
$v_m^2$ the    mean square velocity  (and  $ln\Lambda  \sim$  12); the
N$_*$/N$_s$ ratio  is between  the actual  star  number N$_*$  and the
simulated number of particles N$_s$.  The nature of  the orbit is also
shown  (together with the Galaxy  model, see below), and the effective
peri- and apocenter values.
 Let us note that the concentration chosen are in the low range,
because of computing constraints: large concentrations require a
high spatial resolution, i.e. a large number of particles.

\begin{table*}
\caption[1]{Parameters of the cluster models and orbits}
\begin{tabular}{llllllllllllll}
\\
\hline
\\
Model&Mass&c&$r_c$&$r_h$&$r_t$&W$_0$&n$_0$&t$_{r0}$ & t$_{rh}$ & N$_*$/N$_s$ 
& Orbit & $r_{peri}$ & $r_{apo}$ \\ 
 &10$^5$ M$_\odot$ &&pc&pc& pc &  & pc$^{-3}$ &  Gyr&Gyr& & & kpc & kpc\\ 
\hline \\
m1 &  3.  &1.04&2.8& 7.13  &31 & 6  &1090& 0.82 &  3.0  & 8.2  & Gal-1 + polar orbit&1.4&5.7\\ 
m20&  4.9 &0.89&7.8& 16.2 &60 & 4  & 114 & 10.7 & 15.3 & 13.4 & isolated & -- & --\\ 
m2 &  4.9 &0.89&7.8& 16.2 &60 & 4  & 114 & 10.7 & 15.3 & 13.4 & Gal-1 + disk orbit& 3.1 & 5.0\\ 
m2r &  4.9 &0.89&7.8& 16.2 &60 & 4  & 114 & 10.7 & 15.3 & 13.4 & same +rotation& 3.1 & 5.0 \\ 
m22 &  4.9 &0.89&7.8& 16.2 &60 & 4  & 114 & 10.7 & 15.3 & 13.4 & Gal-2 + disk orbit& 2.8 & 3.2\\ 
m2ret&  4.9&0.89&7.8& 16.2 &60 & 4  & 114 & 10.7 & 15.3 & 13.4 & same+retro- rot.& 2.8 & 3.2 \\ 
m3 &  18. &1.05&9.2& 23.7 &104& 6 & 174 & 12.2 & 44.1 & 48.7 & Gal-1 + disk orbit& 2.0 & 3.4\\ 
m4 &  4.9 &0.89&7.8& 16.2 &60 & 4  & 114 & 10.7 & 15.3 & 13.4 & Gal-2 + disk orbit& 2.2 & 3.2\\ 
m4r&  4.9 &0.89&7.8& 16.2 &60 & 4  & 114 & 10.7 & 15.3 & 13.4 & same +rotation& 2.2 & 3.2\\ 
m5 &  0.9 &1.00&7.2& 17.9 &71 & 5  &  18  & 2.94 &   6.9 & 2.5   & Gal-2 + polar orbit& 17. & 30.\\ 
m6 &  1.2 &1.41&2.0& 10.1 &51 & 10& 870 & 0.08  & 2.5  & 3.4   & Gal-2 + disk orbit& 7.7 &10.\\ 
m7 &  4.9 &0.89&7.8& 16.2 &60 & 4  & 114 & 10.7 & 15.3 & 13.4 & Gal-3 + disk orbit& 2.8 & 3.2\\ 
m8 &  0.9 &1.00&7.2& 17.9 &71 & 5  &  18  & 2.94 &   6.9 & 2.5   & Gal-3 + polar orbit& 21.8 & 30.\\
\hline
\end{tabular}
\label{gct}
\end{table*}

\subsection{Galaxy model}

\begin{figure}
\psfig{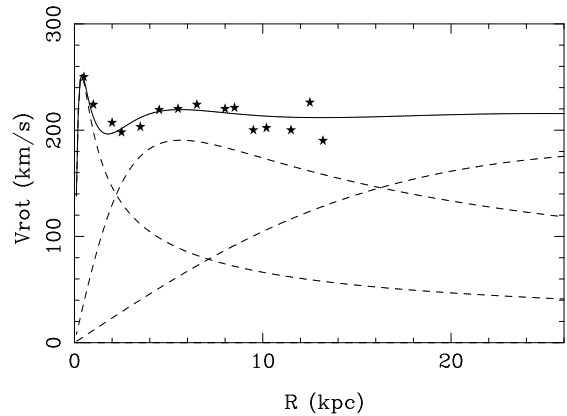}
\caption{Rotation curve  resulting from the  choice of the  three mass
components (parameters Gal-2 displayed  in Table 2), compared with the
data points  for  the Milky Way,  as  summarised by   Fich \& Tremaine
(1991).}
\label{vrot}
\end{figure}

\begin{figure}
\psfig{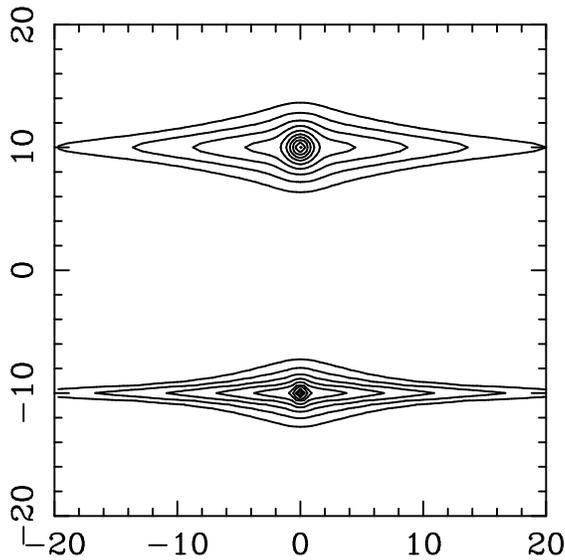}
\caption{Vertical cuts of the visible density  distribution for two of
the models tested for the Galaxy: {\it top} the Gal-1 model ($h_d$ = 1
kpc), and {\it bottom} the Gal-2 model ($h_d$ = 0.4 kpc). The bulge is
included. Scale is in kpc.}
\label{miya}
\end{figure}

We model the potential of the Galaxy by three components,
bulge, disk and dark matter halo. The bulge is a spherical
Plummer law:
$$
\Phi_b(r) = - G M_b (r^2 + a_b^2)^{-1/2}
$$
corresponding to the total mass  $M_b$ and characteristic size  $a_b$
($r$ is the spherical radius);
the disk is  a Miyamoto-Nagai  model,   with mass  $M_d$   and scale
parameters $a_d$ and $h_d$:
$$
\Phi_d(r_c,z) = - G M_d (r_c^2 + (\sqrt{z^2+h_d^2} + a_d )^2 )^{-1/2}
$$
where $r_c$ is the cylindrical radius.
The dark matter halo is added to obtain a flat Galactic rotation curve
$V = V_h$ in the outer parts, i.e.:
$$
\Phi_h(r) = -\frac{1}{2} V_h^2 ln (r^2 +a_h^2)
$$
We tried  some  extreme models, maximum  disk or  not, to  explore all
possibilities for the  disk  mass, that can  give  very different disk
shocking  efficiencies,  for the same rotation  curve.   The first two
models (Gal-1   and Gal-2, see  Tables  \ref{gct} and \ref{galm}) have
spherical  dark matter haloes, but   Gal-2 has  the most  concentrated
disk, so that  the  disk surface density  is  much higher towards  the
center.  Both  have  comparable rotation  curves in  the Galaxy plane.
The thickness of  Gal-2  is equivalent to  that  of an  exponential of
scale-height 300pc, still somewhat higher  than the recent estimations
of 250pc (Haywood et al.  1997) for  the Milky Way.   A summary of all
parameters used  is displayed in Table  \ref{galm},  and the resulting
rotation curve of Gal-2 is compared in Fig.~\ref{vrot} to the observed
Milky  Way rotation curve.  The density distribution  in the first two
models of the galaxy is displayed through  density contours in central
cuts in Fig.~\ref{miya}.

The third model Gal-3  has been run to  test an extreme flattening  of
the dark  matter halo component.  For that  reason, we have not chosen
to  generalise  the  above     spherical   logarithmic potential    to
isopotentials on concentric ellipsoids,  since at large flattening, it
corresponds to unphysical density distributions.  In this third model,
the visible components  are of the  same  form as previously, and  the
dark halo mass   density is given   by  a pseudo-isothermal  ellipsoid
(Sackett \& Sparke 1990):
$$ 
\rho_h(R,z)=\rho_0 \Bigl[1+{{(r_c^2+z^2/q^2)}\over{a_h^2}}\Bigr]^{-1} 
$$ 
where $\rho_0$ is the central density,  $a_h$ the core radius, and $q$
is the axial ratio of the isodensity curves, which vary from spherical
($q$ = 1) to flattened ellipsoids ($q < $ 1).

The potential corresponding to this density is:
$$
\Phi_h(r_c,z) = -2 \pi G q  \rho_0 a_h^2 \int_0^{1/q}
{ {ln \Bigl[ 1 +\frac{x^2}{a_h^2} (\frac{r_c^2}{x^2\epsilon^2 +1}+z^2) 
\Bigr] dx}  \over {x^2\epsilon^2 +1}  }
$$ 
where $\epsilon^2   = 1-q^2$. Forces  can be  derived analytically, in
cylindrical  coordinates  (Sackett  et al.  1994)  and  in ellipsoidal
coordinates (de  Zeeuw \& Pfenniger 1988).   This model also  gives an
asymptotically flat rotation  curve ($V_h$),  that  we will  refer to,
instead of the central density $\rho_0$:
$$
V_h^2 = 4 \pi G \rho_0 a_h^2 q Arccos(q) / \epsilon
$$
and the mass included in the ellipsoid of axes $a$ and $aq$ is:
$$
M_h = { {V_h^2 a_h\epsilon} \over {G Arccos(q) } } 
\Bigl[ \frac{a}{a_h}-Arctan\frac{a}{a_h} \Bigr]
$$
The model used (Gal-3, parameters in  Table \ref{galm}) is chosen with
an extreme flattening of $q=0.2$ to probe the effect.

\begin{table}
\caption[1]{Parameters of the mass models for the Galaxy}
\begin{tabular}{llll}
\\
\hline
\\
Model Gal-1  &  Bulge & Disk & Dark Halo (log)\\ 
\hline \\
Radial Scale  (kpc) &  0.5 & 6.0 & 10. \\ 
Mass (10$^9$ M$_\odot$) &  19.8 & 72. & 190. ($V_h$ km/s) \\ 
Vertical Scale (kpc) &  -- & 1.0& -- \\ 
\hline
\\
Model Gal-2  &  Bulge & Disk & Dark Halo (log)\\ 
\hline \\
Radial Scale  (kpc) &  0.25 & 3.5 & 20. \\ 
Mass (10$^9$ M$_\odot$) &  9.9 & 72. & 220. ($V_h$ km/s) \\ 
Vertical Scale (kpc) &  -- & 0.4& -- \\ 
\hline
\\
Model Gal-3  &  Bulge & Disk & Dark Halo (isoth)\\ 
\hline \\
Radial Scale  (kpc) &  0.25 & 3.5 & 10. \\ 
Mass (10$^9$ M$_\odot$) &  9.9 & 72. & 220. ($V_h$ km/s) \\ 
Vertical Scale (kpc) &  -- & 0.4& $q$ = 0.2 \\ 
\hline
\end{tabular}
\label{galm}
\end{table}

\begin{figure}
\psfig{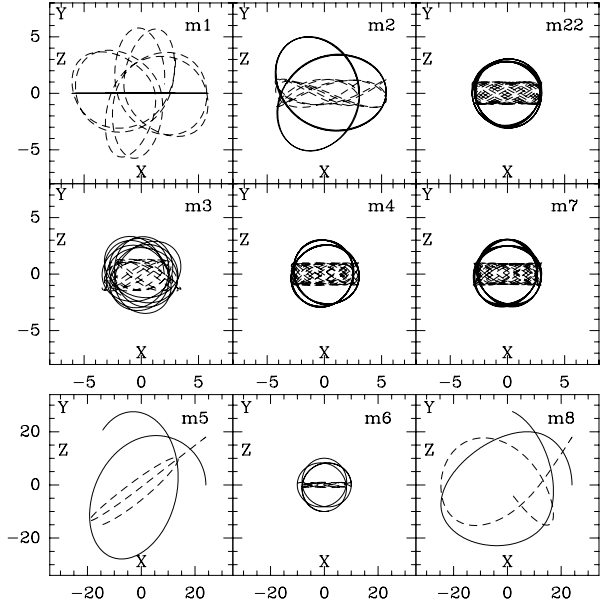}
\caption{ A sample of  orbits used in the  simulations, polar and disk
orbits,  projected in the xy (solid  lines)  and xz (dash) planes. The
scale is in kpc. The orbits are integrated for one Gyr. }
\label{orbits}
\end{figure}

In  these potentials, essentially two   kinds of orbits were selected;
the  nearly polar  orbits, and the   "disk" orbits, where  the cluster
crosses the disk very frequently (see Fig.~\ref{orbits}).

\section{Results and Discussion}

\subsection{Relaxation}

\begin{figure}
\psfig{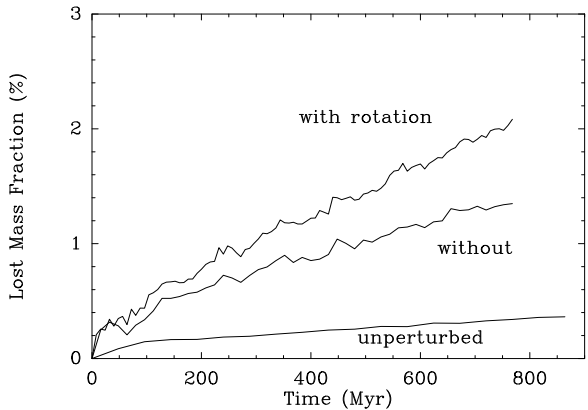}
\caption{ Fraction of unbound mass as a function of time, for run
m20 (isolated cluster), m2 (without rotation), and m2r with rotation and
the same orbit.}
\label{lostm2}
\end{figure}

First  we want to estimate  the degree  of  relaxation provided by our
N-body scheme.   The relaxation in actual star  clusters is due to the
granularity   of the potential created    by   the finite number    of
stars. This granularity is larger with  a reduced number of particles,
so  the relaxation is accelerated  by simulating a number of particles
inferior to the real one. On  the contrary, the two-body relaxation is
reduced by  the softening  of the potential  at  small scale.  The two
effects are somewhat  compensating in a complex  way, and  we can only
estimate the resulting rate  of relaxation numerically.  In  any case,
even if the resulting relaxation time is comparable to the actual one,
the  effects will not be  the same in  terms of spatial frequency.  In
Table \ref{gct}    are indicated the real  relaxation   times, and the
N$_*$/N$_s$ ratio between  the  actual   star  number N$_*$  and   the
simulated number of  particles N$_s$; without softening, the  expected
relaxation  time in  the simulations is  shorter than  t$_{rh}$ by the
factor N$_*$/N$_s$. We have run the  model m20 isolated to measure the
rate of evaporation precisely due to the relaxation. The measured mass
loss after 0.86  Gyr  is  0.3\%  (see  Fig.~\ref{lostm2}).  Since  the
expected loss by evaporation alone is of the order of 4\% per t$_{rh}$
(H\'enon 1961, Gnedin \& Ostriker 1997),  we infer that the equivalent
t$_{rh}$ is 12  Gyr, not so  far from the actual  one (15.3  Gyr).  At
least, rapid relaxation in the simulations are not perturbing too much
the dynamical evolution we want to follow here.

\subsection{Mass loss}

The computation of the amount of unbound particles at a given epoch is
delicate.  The concept  of a   tidal  radius separating the  bound and
unbound stars is clear only when the globular cluster is embedded in a
steady potential, which is the case for an ideal circular orbit in the
galactic  plane for instance, without any  disk crossing (e.g. Spitzer
1987). Globular  clusters are observed with   a limiting radius, where
the  surface density drops (e.g.    Freeman \& Norris, 1981). Attempts
have been made through    cluster modelling to relate   this  observed
cut-off to the tidal radius, as defined by King (1962).  Keenan (1981)
found that the   limiting radius was  close to   the  tidal radius  at
pericenter,  while it was   interpreted as the   local tidal radius by
Innanen    et al.  (1983).   Since  globular  clusters  orbits are not
precisely known, and also the limiting radii  are only determined with
large uncertainties, the  situation remains unclear. However, for very
excentric orbits,  the   tidal radius varies considerably   along  the
orbit, and  not all particles which become  unbound  at pericenter are
still so at apocenter.

\begin{figure}
\psfig{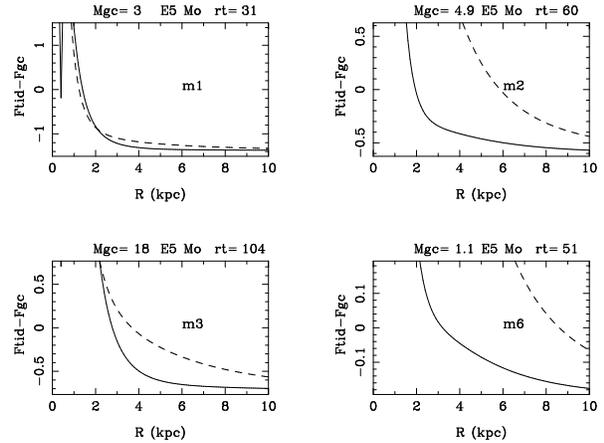}
\caption{  Differences  between  the  intensity  of   the tidal  force
($F_{tid}$)  felt  by stars  at the surface   of  the globular cluster
(r$_t$), and the proper attracting force  of the cluster ($F_{gc}$) as
a  function of galactic  radius,  for the  various cluster models. The
full line corresponds to the radial tidal  force, and the dash line to
the vertical one (at $z=0$). When the curves cross the horizontal axis
(i.e.  the differences  vanish),  they define  the  minimum pericenter
distance  corresponding  to the  limiting  radius of  the cluster. The
corresponding runs are indicated (m1,m2,m3,m6).}
\label{tidf}
\end{figure}

Moreover, the strongest tidal force is in fact the force perpendicular
to the plane, felt by particles at disk crossing, everywhere except in
the bulge (see Fig.~\ref{tidf}).  Although  the vertical forces always
correspond  to a compression, they  yet give energy  to the particles,
and  trigger a rebound or oscillation,  produce a  vertical tidal tail
and can unbind the particles. The vertical force gradient is therefore
dominant  in the Galaxy,  except for the  bulge and the remote regions
dominated by the halo.

For  each  cluster model,  we   have taken into  account  the required
galactic force gradient to explain  its limiting radius, to choose its
corresponding orbit.  This ensures that the cluster is not launched in
a completely un-realistic manner, with its limiting radius much larger
(or much  smaller)  than its  tidal   radius, in which   case it would
quickly have  lost  a large fraction  of its  mass (or   it would have
relaxed in a long time to another limiting radius, without mass loss).
We therefore hope to reach quickly a quasi-steady state, and determine
the corresponding tidal tail, while estimating the mass loss rate.

In determining  the mass loss  at a given  epoch, we take into account
the  dynamical evolution of the   cluster (in concentration and mass),
but keep the force gradients of the galaxy constant, and equal to that
at pericenter.  We  solve at each  epoch  the equation for   the tidal
radius, which slightly  decreases as evolution  proceeds, and consider
stars unbound  when their relative  energy is positive, i.e. $\epsilon
<$0, where:
$$
\epsilon =\Phi(r_t) - \Phi(r) - \frac{1}{2} v^2
$$

\begin{figure}
\psfig{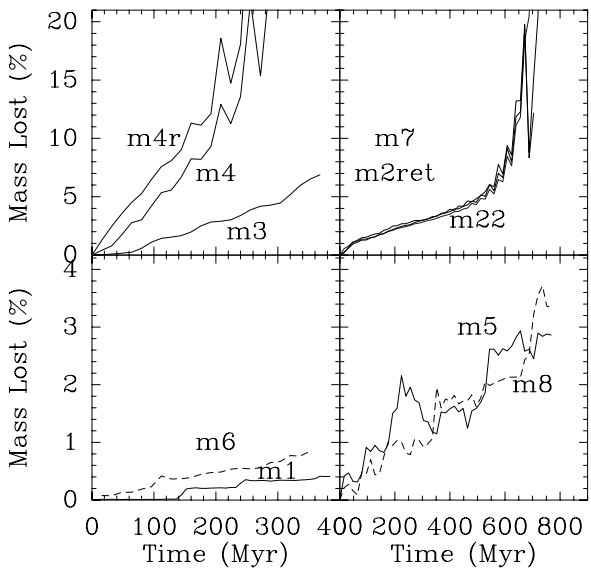}
\caption{  Fraction of unbound  mass as  a function  of time, for  the
different runs.    Note that the retrograde  rotation  (m2ret)  of the
globular cluster has  no effect, nor the flattening  of the dark  halo
for inner orbits (m7) since the curves are then almost coinciding with
the standard run (m22).  The  flattening of  the dark  halo,  however,
plays a role for outer orbits (cf m8/m5). }
\label{lost}
\end{figure}

The lost-mass fraction for the model m2, m2r  (with rotation), m20 the
comparison isolated cluster, is  plotted in Fig.~\ref{lostm2}, and for
all other  runs  in Fig.~\ref{lost}.  For  almost   all the runs,  the
gravitational  shocks are more efficient  than   the evaporation by  a
factor 1 to 100.  Only  the run m1 which  has a very short  relaxation
time (3    Gyr)  has an  evaporation time     scale   lower than   the
gravitational shock   time-scale   in the  thick disk  Galaxy  model
Gal-1. Moreover the relatively low concentration of the clusters simulated 
here locates them in the most sensitive  branch 
of the curve of the evaporation time versus concentration shown  by
Gnedin \& Ostriker (1997),  namely $T_{evap}/T_{rh}$ varies between 20
and  30 for our  set of simulations. The   disk/bulge shocking is very
efficient to destroy in a very rapid phase  the cluster m22 because of
the thinner disk of  the Gal-2 model.  In the  similar case of run m3,
in spite of the Gal-1  model, the mass loss  is important, because  of
the  large size of  the  cluster. Observational  studies of mass  loss
combined with  cluster parameters and   reliable orbits will constrain
strongly the disk/bulge parameters.

\subsection{Influence of rotation}
Merritt et al. (1997) have studied in details the rotation of $\omega$
Centauri, from radial  velocity data of about  500 stars.  The cluster
is in axisymmetric, non cylindrical rotation,  with a peak of 7.9 km/s
(at  11~pc from    the center, i.e.   0.15  $r_t$).   Their rotational
velocity      corresponds  well   with      our  rotating  model  (see
Fig.~\ref{roprof}).  Drukier et  al.  (1998) also analysed in  details
the kinematics of  230 stars  in  M15, and   found that  a model  with
rotation is  marginally favored   over  one without  rotation.    They
discover that the velocity  dispersion increases slightly in the outer
parts, indicating the possible heating by the galactic tides.

Keenan   \& Innanen  (1975)  have  shown that   clusters rotating in a
retrograde sense are more stable in the tidal field of the Galaxy than
direct   rotating or  non-rotating  clusters. They   also followed the
orbits  of escaped stars, and  found that they  can stay  in the tidal
tail  for a large part of  the  cluster orbit.  However,  they use the
three-body integration scheme, taking  no account of self-gravity  and
relaxation.

We have run several   models with and  without  rotation, to test  the
effect on the mass loss of globular clusters  in the Galaxy.  When the
rotation of the  cluster is in  the direct  sense with  respect to its
orbit, the mass loss appears  higher than for models without rotation:
it     is the  case  for      m2/m2r (Fig.~\ref{lostm2}), and   m4/m4r
(Fig.~\ref{lost}).  However, the  difference  is negligible  when  the
rotation of the cluster is in  the retrograde direction (cf m22/m2ret,
Fig.~\ref{lost}).  This phenomenon has been seen and explained in many
circumstances,  including galaxy interactions.  It comes from the fact
that particles rotating  in the direct  sense in  the cluster resonate
more with the   galaxy  potential, while  the relative   velocities of
cluster stars  and the Galaxy are higher  in the  retrograde case, and
the perturbation  is  then more  averaged out.  As a consequence,  the
directly rotating  clusters are  disrupted  earlier, and  there should
remain today an  excess of retrograde clusters.  This is difficult  to
check statistically (it  can be  noted that  the  angular momentum  of
$\omega$ Cen  is anti-parallel to  that  of the Galaxy, and  therefore
compatible with predictions).

\subsection{Influence of the cluster concentration and
of its orbit}

\begin{figure}
\psfig{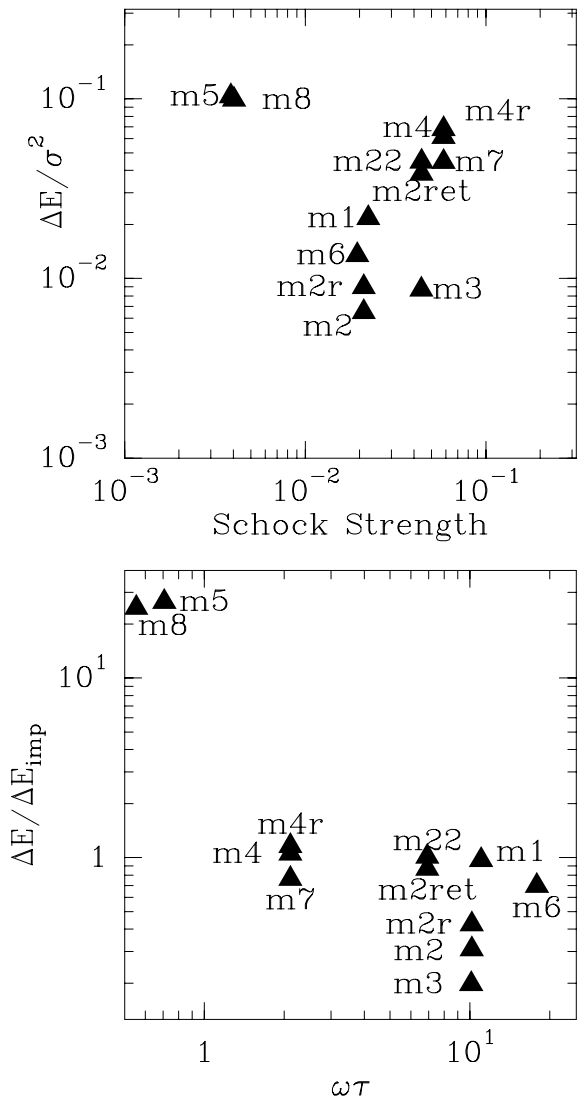}
\caption{  Upper  panel:  total   energy   change  per disk   crossing
($\Delta$E) for the different  runs, as a  function of  shock strength
($\Delta$E$_{imp}$), defined from  the  expected heating  rate  in the
impulse approximation (both  normalised  to the  internal   velocity
dispersion   $\sigma_0$,  cf  text).     Lower  panel:  ratio  $\Delta$E
/$\Delta$E$_{imp}$   as    a function   of   the  adiabatic  parameter
$\omega\tau$. }
\label{choc}
\end{figure}

Fig.~\ref{lostm2} and   \ref{lost} demonstrate clearly the   effect of
disk  shocking: between 1 and 20  kpc in radius, the z-acceleration of
the disk close to  the plane decreases by  100, and between the models
of Gal-1 and Gal-2 (maximum disk) the acceleration  is multiplied by 4
near the  center. This explains  that the run  m4 and  m4r lead to the
destruction of the cluster, while only a slight  mass loss is observed
for m2 and m2r.   The more concentrated clusters  (m1 and m6) are also
much  less  affected, even at low  pericenter,  and with maximum disk.
For polar orbits, the mass loss curves show clear flat stages, between
steps corresponding to each disk crossing (m1, m5,  m8). There is even
some bouncing effect: the  particles unbound to the  cluster by a disk
shocking  can be  re-captured during a  more  quiet phase (m5).   Less
steep ondulations  are observed  for the disk   orbits, at  each  disk
crossing (m3, m4).  To better quantify the  disk shocking effects, and
to compare it to the heating expected  from the impulse approximation,
the  heating  per  disk crossing, at   the beginning  of  the  runs is
estimated  in Fig.~\ref{choc}.  The  shock  strength is defined as the
shock heating  per unit  mass in  the impulsive approximation  $\Delta
E_{imp}$, normalized  to  the internal   velocity dispersion  of   the
cluster, i.e. (Spitzer 1987):
$$
\Delta E_{imp} = 2 z^2 g_m^2 / V^2
$$ 
where  $g_m$ is  the  maximum z-acceleration due   to the disk, $z^2 =
r_h^2/3$ characterizes the typical size of the cluster, and $V$ is the
z-velocity at which it crosses the  disk.  To estimate the heating per
unit  mass and  per shock in  the simulations,  we computed the total
energy of the   cluster in the   first  100-200 Myr of  their  orbits,
averaging the  heating over 3-7 shocks (except  for the polar runs m5
and m8, where we considered  only one disk  crossing, since the second
occurs after 400 Myr).

Fig.~\ref{choc} (upper panel) shows that there  is a rough correlation
between   the  two   quantities,    $\Delta  E/\sigma_0^2$  and  $\Delta
E_{imp}/\sigma_0^2$, expected   if  the  impulsive   approximation    is
applicable. The polar   runs   m5 and  m8  are  however  outside  this
correlation, since    their interaction with   the Galaxy   cannot  be
approximated at all by shocks at disk crossing, given the weakness of
the disk they encountered on  their orbits.  All  other runs display a
heating   rate below  that from   the impulse  approximation, which is
expected  if an  adiabatic    correction is  added.   We  define  this
correction  as  usual from  the  product  of  the   internal frequency
$\omega$  by  the  time-scale   of    the perturbation  $\tau    =H/V$
(crossing-time of the disk height $H$), and $\omega$ is related to the
central density $\rho_0$ such that 

$$\omega^2 = G 4/3 \pi \rho_0$$

The ratio of  the observed heating  rate  to the expected one  for the
impulse  approximation is plotted versus   this adiabatic parameter in
Fig.~\ref{choc} (lower   panel).  It shows   clearly  that a  negative
adiabatic  correction should be added.  This correction appears not as
steep   as an  exponential,   as  predicted   by Spitzer  (1987),  and
corresponds  more  to  a power-law  as  found  by   Gnedin \& Ostriker
(1999). A  more detailed estimate   of this correction  should involve
individual stellar orbits in  the cluster, and  is beyond the scope of
this study.

\subsection{Influence of the dark matter flattening}

Two models were  run with an extreme  flattening for the dark halo (an
axis ratio of $q$ = 0.2).  In the first one  (m7), the small radius of
the orbit places it  in a region  where the dark halo contribution  to
the potential is  not large, and  the difference with a spherical halo
is  not  significant.  For the  second  one (m8)  the  orbit peri- and
apocenters are both  in a region dominated  by the dark  halo, and the
orbit is polar,  so    that the  halo  flattening  effect    should be
maximum. The difference with the comparable  run (m5) with a spherical
halo is easily seen (Fig.~\ref{lost}), but  the overall mass loss rate
are comparable.  This means that the  tidal shocks  at the crossing of
the flat    halo  are   not  strong  enough    to compete   with  disk
shocking. Also, there  is less mass in the  flattened halo model, with
respect to the spherical model, for  the same rotation curve. Inside R
=  26 kpc, the dark  halo mass is 1.1  and 1.8 10$^{11}$ M$_\odot$ for
the flattened and spherical models,  respectively. Near the plane, the
force per unit mass  can be approximated as $K_z$z,  and the  value of
$K_z  = -{{\partial^2 \Phi} \over  {\partial  z^2}}$ is about 5  times
higher  for the flattened  than for the  spherical  dark halo. But the
disk $K_z$ is always larger than that of the flattened dark halo; they
begin to equalize only at radii of 30  kpc. That explains why the halo
shocking is  not  significant.  The  globular  cluster dynamics is not
useful  to constrain the flattening of   dark haloes, nevertheless the
halo geometry will affect the destruction rate of the remote clusters.

\begin{figure}
\psfig{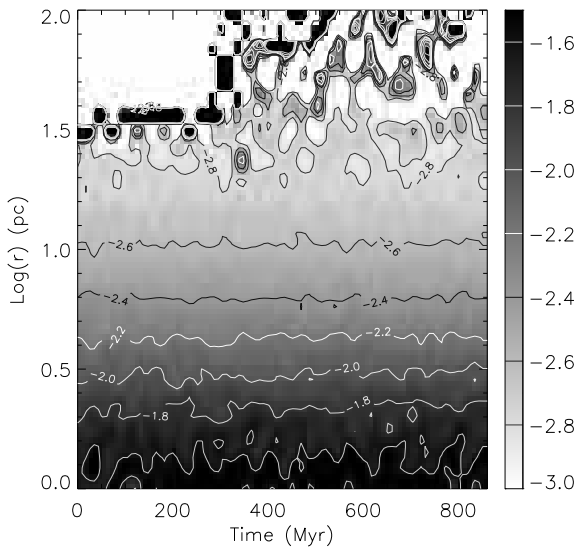}
\caption{ Slope  of the mass distribution  power-law  as a function of
time and radius in the cluster,  for run m1.  Since the low-mass stars
are  concentrated   in the  outer     parts, they  are  preferentially
stripped. }
\label{slopem1}
\end{figure}

\subsection{Mass segregation}

\begin{figure}
\psfig{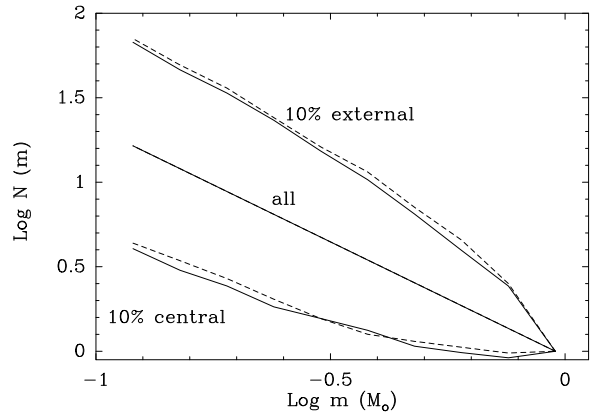}
\caption{  Mass distribution function  at different radii for model m2
(without rotation).  The  mass spectrum inside   the radius containing
10\% of the  mass is compared with that  outside the radius containing
90\%  of the  mass.  Dash lines correspond   to  the beginning of  the
simulation, and full lines at the end. }
\label{massf}
\end{figure}

It  is well  known that  the mass  distribution function for  low-mass
stars in  an old globular cluster  is very close to  the IMF;  this is
specially  true  at the   radius $r_h$,  region which   is more robust
against mass segregation and   tidal stripping (Vesperini, 1997),  the
two processes  somewhat compensating  at this  radius to  preserve the
IMF.  In  the multi-mass King-Michie  model that we adopted as initial
conditions,  the mass  function  varies  with radius,  as sketched  in
Fig.~\ref{massf}.   It  is  also expected  that   the tidal  stripping
increases the mass segregation, in acting together with the relaxation
of the  central parts. Since  the envelope is preferentially populated
by low-mass stars,  they are  more stripped  than the high-mass  ones.
This is  shown in Fig.~\ref{slopem1}, where the  value of the slope of
the mass  function is  plotted  (in gray-scale and  in contours)  as a
function of time and radius in the cluster.  Core  relaxation is a way
to replenish  the low-mass stars content of   the envelope. Recent HST
results  have discovered that some  globular clusters have indeed mass
functions  depleted in low-mass stars (Sosin  \& King  1997, Piotto et
al. 1997); the depleted clusters are very likely candidates for recent
tidal shocking, from   their  presumed present galactic  position  and
orbit  (Dauphole  et al. 1996).  

The evolution through  its dynamical  life of the  mass  function of a
globular cluster has been widely studied, in the  goal of deriving the
IMF slope at low mass of the galactic halo itself, which has important
implications for the nature of dark matter.  Gnedin \& Ostriker (1997)
showed that about 75\% of the present globular cluster population will
be destroyed in   a Hubble time,   and therefore that the majority  of
initial clusters is now  destroyed and forms a  large fraction of  the
stellar halo and bulge.  Vesperini  (1998), using an analytical scheme
for the destruction   processes, estimated the   initial population of
globular clusters  to be about  300  clusters and the  contribution of
disrupted clusters to the  halo would be about  $5.5~10^7 M_\odot$, of
the same order as the stellar mass in  the halo (Binney \& Merrifield,
1998).    This means that the    remaining clusters must have  evolved
considerably, and in particular   their mass function.  Capaccioli  et
al. (1993)  have observed that the  mass function slope  is correlated
with the position of  the globular cluster in  the galactic plane, and
in particular  with its height. The  mass function is steeper at large
distance. They show through analytical   and N-body calculations  that
this could  be due to disk shocking,  that flattens the mass function.
Johnston  et al. (1999)  have studied  through  N-body simulations the
mass loss rates   in mass-segregated systems.   They confirm  that the
mass function  is  considerably flattened during  tidal evolution; the
slope x can fall  from 1.35 to almost 0  for the  more tide-vulnerable
cases (small clusters in disk orbits).

\subsection{Tidal tails}

\begin{figure}
\psfig{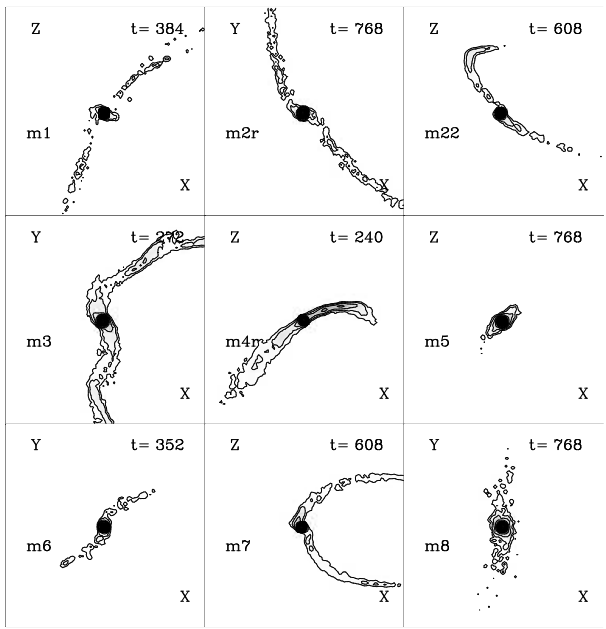}
\caption{ Contours of the projected density  of particles (either onto
xy, or xz planes) in logarithmic  scale, with various clipping to show
the tidal tails for the 9  different models indicated.  Each square is
3.6 kpc in size.  The epoch is indicated in Myr.  }
\label{figtail}
\end{figure}

Once the  particles are unbound, they  slowly drift along the globular
cluster path when they were launched, and form a huge tidal tail. They
can   still  form a recognizable  features  well  outside  the cluster
envelope,  and  hundreds   of pc away,   as  is  observed on   the sky
(Grillmair et  al. 1995,  Leon et al.  1999).   The tidal tails  for 9
different models are    displayed in Fig.~\ref{figtail},  at  the same
spatial scale.

The unbound particles,  and therefore the tails,  are a tracer  of the
cluster orbit. The tails look asymmetrical, with a heavier tail on one
side with respect to  the other, but this  is a projection effect, due
to the  particular shapes of  the  tails. There are  sometimes special
wiggles at the basis  of the tails,  in the cluster envelope, that are
not due to the rotation  of the clusters,  since they are seen  around
non-rotating clusters  as well   (Fig.~\ref{figtail}). These will   be
interpreted in Section 3.8.
 
\begin{figure*}
\psfig{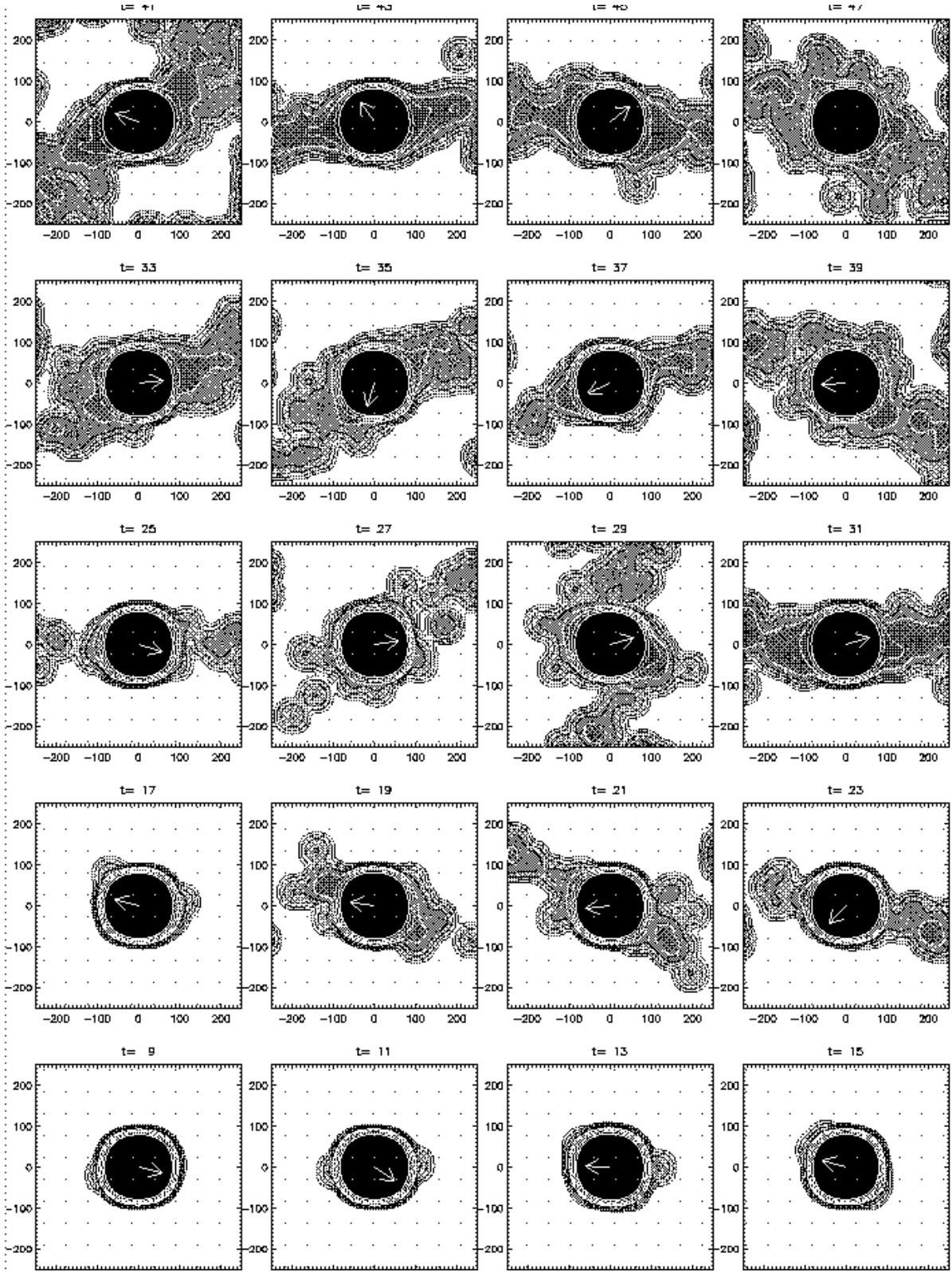}
\caption{ Tidal tails mapped at different epochs with  the  wavelet
algorithm, for the model m2; the direction of the galactic center  
is indicated by the  arrow. The maps are projected in the (X,Z) axes,
OZ being perpendicular to the galactic plane.}
\label{m2wave}
\end{figure*}

Analysis  of   the faint tails  is  best  performed  with  the wavelet
decomposition,   that     can  achieve   multi-resolution,     as   in
Fig.~\ref{m2wave}.  The tidal tails can contain  a few percents of the
mass      of   the   cluster.      Fig.~\ref{figtail}, \ref{m2wave}
  and   especially
Fig.~\ref{torsion}  show the clumpy  structure in the tidal tails. The
denser unbound  clumps are the tracers of  the strongest  phase of the
gravitational shocks: the two  symmetrical counterparts are visible on
each side of the tails. 
 Although these clumps are not bound, but transient caustics,
the structure of the tail remains clumpy all over the simulations, 
even if it is not the same stars in the same clumps. We have followed
in the simulations a group of particules that formed a clump at 
a given epoch: the group stays together for some time, until
the packet of particules moves away from the cluster, then another clump
is formed by a new tidal shock. It takes more than 800 Myr for a clump to
disperse. A typical clump can contain 0.5\% of the cluster stars.
Some observed clusters (Leon et al. 1999) show
evidence for such features  in their tidal tails  (e.g. NGC  5264, Pal
12). These overdensities in tidal tails are related to the "streamers"
or  moving groups  in  the halo   (Aguilar,  1997). These  symmetrical
features have been  detected as well for  open clusters oscillating in
the galactic plane (Bergond et al. 1999).

How close tails  trace the globular cluster orbit  can be clearly seen
on   Fig.~\ref{tailm4}, showing a  large-scale  view  of the  globular
cluster of the run m4, in  a disk-like orbit. This cluster experienced
strong  disk shocks,   due to the  thin-disk  model  chosen,  and was
disrupted in 0.5 Gyr.

\begin{figure}
\psfig{figure=8931.f15,bbllx=1cm,bblly=1cm,bburx=21cm,bbury=135mm,width=8cm,angle=-90}
\caption{  Contours of the projected density  of  particles in the run
m4, in the logarithmic scale, showing the fate of the globular cluster
in the  thin-disk galaxy model.  The squares shown  are  8 kpc across,
centered on the galactic center.  The huge tidal tails trace perfectly
the  cluster orbit,  even the  loops in the   x-z projection. Time  is
indicated in Myr.  }
\label{tailm4}
\end{figure}

\begin{figure}
\psfig{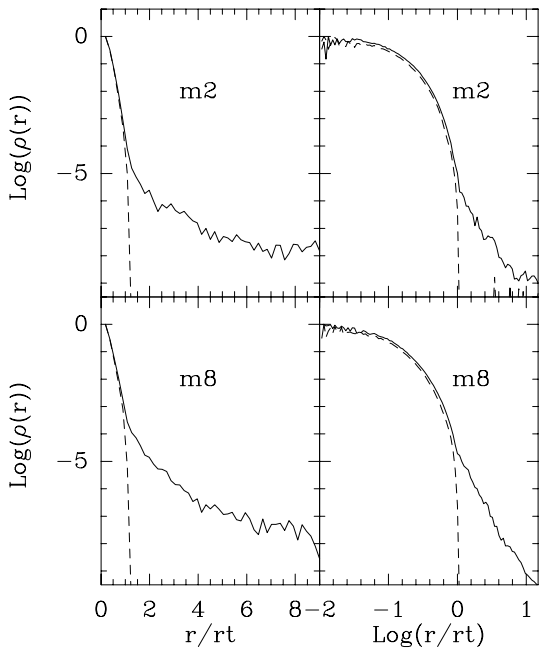}
\caption{ Density profiles  of two of the  runs (m2 and  m8), at t=768
Myr, both as a  function of radius (left)  or radius log  (right). The
dash line is  the density at t=16 Myr  for comparison. Note  the clear
change  of slope  at the  tidal   radius $r_t$. The  slope in  volumic
density of the tail is around --4 (and in surface density, --3).}
\label{slope}
\end{figure}

Unbound particles, outside the tidal radius of the cluster, spread out
in density  like   a power law,   with   a slope  in average   of --4.
(Fig.~\ref{slope}).  This kind of  behaviour has been found  for tidal
extensions  in  numerical  simulations  of  interacting  galaxies  (cf
Aguilar \&  White 1986), and is  that expected of  an unbound cloud of
particles in  an 1/r potential,  with a  continuous  spread in  energy
(with an almost constant probability, N(E), since then N(r)dr $\propto
\rho r^2 dr \propto$ N(E)dE $\propto  r^{-2}$) starting from just zero
relative energy, since at large  distance, the globular cluster can be
considered as a point mass, with 1/r potential.  It does not depend on
the  source of the  mass loss,  and   not significantly on  the Galaxy
potential either, since the latter gradients develop on larger scales.

The corresponding surface density around the cluster falls as $r^{-3}$
or  steeper, which  is far from   the predictions  of  Johnston et al.
(1999)  for independent tidal  debris, free to move  and spread in the
Galaxy  potential: their  expected slope  of  surface density is  --1.
Although these particles are in majority  unbound to the cluster (have
positive energy, as defined in Section 3.2), they cannot be considered
as  completely free,  but still under    the influence of the  cluster
potential; in particular, the closest ones can still be re-captured by
the cluster.   The   discrepancy with the observations   (Grillmair et
al. 1995, Leon et al. 1999) where most of the slopes are around --1 is
likely the consequence of noisy background-foreground subtraction.

\subsection{Flattening}

\begin{figure}
\psfig{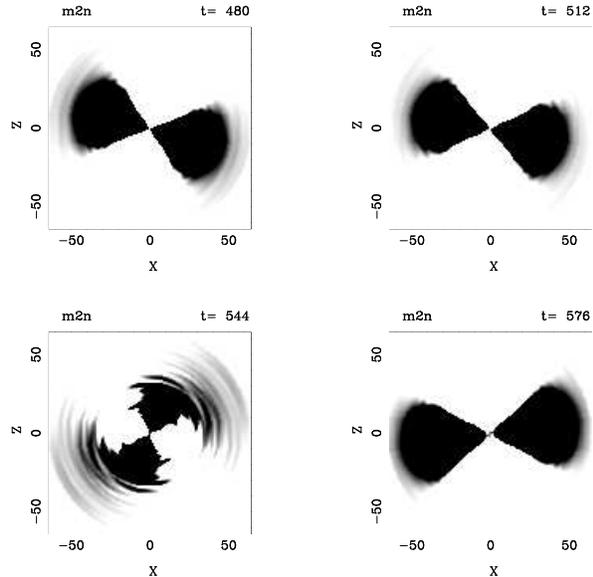}
\caption{ Fourier  components (even harmonics  $m=$2, 4 and  6) of the
surface density  projected  into the x-z  plane for  the model m22 (no
rotation). The scale is in pc, and time is in Myr.}
\label{m2four}
\end{figure}

At smaller  scale, closer  to  the tidal   radius,  the just  escaping
particles  are  in general oriented perpendicular   to the plane, just
after a disk crossing. Monitoring the flattening of the cluster can be
done through a Fourier analysis. Fig.~\ref{m2four} shows the resulting
of even harmonics ($m=$ 2, 4, and 6) of  the surface density projected
in the plane, or  perpendicular to the  plane. When the cluster has no
rotation, periodic compression  of the cluster at  its crossing of the
plane, and    subsequent relaxation,   corresponding sometime    in an
extension of   the cluster in  the  vertical direction  is easily seen
through the orientation of density maxima.

\begin{figure}
\psfig{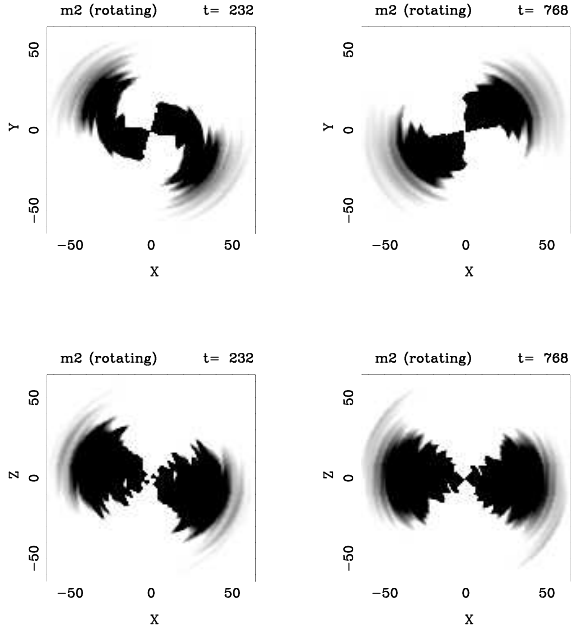}
\caption{ Fourier components (even  harmonics $m=$2, 4  and 6) of  the
surface density projected  into the x-z plane,  and x-y  plane for the
model m2r (with rotation). }
\label{rom2four}
\end{figure}

In the case  of  a rotating cluster  (m2r),  the flattening in  the xz
plane is dominated by the rotation, although it varies slightly at the
disk crossings. In the  equatorial plane (xy), there  is also an $m=2$
perturbation,   which   appears to tumble   in  the  sense of rotation
(Fig.~\ref{rom2four}).

\begin{figure}
\psfig{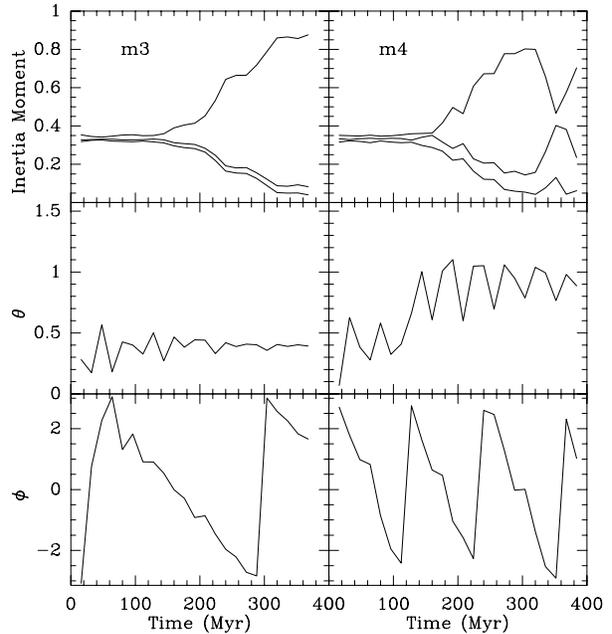}
\caption{ The three eigen values of the inertia  tensor, as a function
of time (top), together with the ($\theta$, $\phi$) orientation of the
major axis  in spherical coordinates  (middle and bottom,  in radians)
for the model m3 (left) and for the model m4 (right). }
\label{gctriax}
\end{figure}

To better characterize the deformation of the globular cluster and the
direction  of  the  flattening,  we have  computed the  (3x3)  inertia
momentum tensor
$$
I_{xy} = { {\Sigma_n m(n) x y/r^2 } \over {\Sigma_n m(n)}}
$$
with all combinations (x,y,z) taken  into account. This computation is
carried out on the  globular  cluster and  its immediate  envelope and
avoids the tidal tails.  The diagonalisation  of this matrix gives the
three eigen values plotted in Fig.~\ref{gctriax} and the eigen vectors
give   the orientation  of the  major  axis  ($\theta$  and $\phi$  in
spherical coordinates). In most of the models the eigen values are all
almost equal to 1/3 (no large perturbations  with respect to spherical
shape). In the most  perturbed cases, it  is  possible to see  a clear
prolate shape (the two  almost equal axes are  the smallest ones) with
the  major  axis,  located  at  an   almost constant  angle   with the
z-direction, and  precessing around this  z-axis (perpendicular to the
plane) with  retrograde sense (see Fig.~\ref{gctriax}).  Some bouncing
effects can also be  noted, for example   in model m4.  The period  of
precession does not depend on the nature of the orbit, and has nothing
to do with the time-scale between two crossings of the plane (it is in
general  much larger).  On the  contrary, the  periods are similar for
runs with the same initial state for the clusters.  We interprete this
period as an eigen value of the cluster itself: once a perturbation is
excited, it develops with these proper  frequencies (see e.g. Prugniel
\& Combes 1992).

\begin{figure}
\psfig{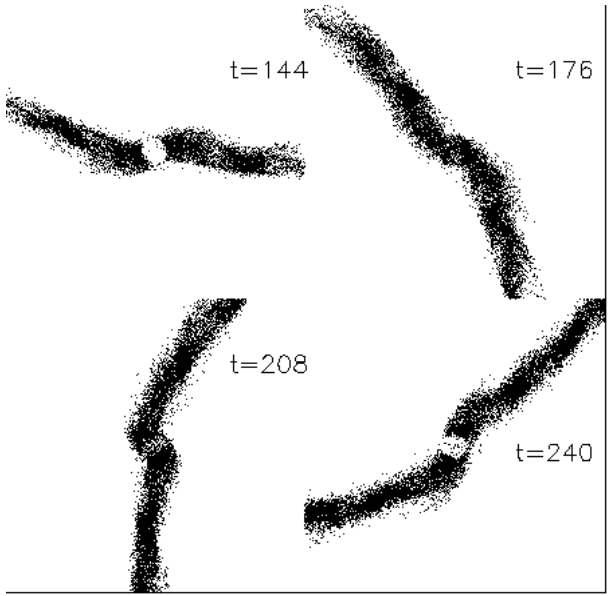}
\caption{ Plot of the particles just outside the tidal radius, at four
epochs of the  run m4 (in Myr). The  particles inside the tidal radius
of  85pc  have  not been plotted,  for   clarity.   The  particles are
projected on the x-y galaxy plane. }
\label{torsion}
\end{figure}

It is  interesting to remark  that  this prolate  shape  taken by  the
globular clusters due  to the tidal  forces orients the mass loss just
at the  beginning  of the tidal  tails:  unbound stars  preferentially
escape in  the direction of  the major  axis,  which gives the crooked
shape of the   global tail, that follows  at  large scale the  cluster
orbit (see Fig.~\ref{torsion}).

\section {Conclusions}

We have carried  out  about a dozen N-body   simulations of the  tidal
interactions  between a globular cluster and  the Galaxy,  in order to
characterize   the  perturbations  experienced at    disk crossing, to
determine the geometry and density distribution of the tidal tails and
quantify the mass loss as a function of  cluster properties and Galaxy
models. Our main conclusions can be summarised as follows:
\begin{itemize}
\item All runs  show that the  clusters are always surrounded by tidal
tails  and  debris,  even those  that  had   only a  very  slight mass
loss. These  unbound  particles distribute  in volumic  density like a
power-law as a function of radius, with a slope around --4. This slope
is    much  steeper    than    in   the    observations  where     the
background-foreground contamination dominates at very large scale.
\item These tails are preferentially composed of low mass stars, since
they are coming  from the external radii of  the cluster; due  to mass
segregation built  up  by  two-body  relaxation,  the  external  radii
preferentially gather the low mass stars.
\item The mass loss is enhanced for a cluster  in direct rotation with
respect to its orbit. No effect is seen for retrograde rotation.
\item For sufficiently high  and rapid mass  loss, the cluster takes a
prolate shape, whose major axis precess around the z-axis.
\item When the tidal tail is very long (high mass loss) it follows the
cluster orbit: the  observation of the tail  geometry is thus a way to
deduce cluster   orbits.   Stars   are not  distributed  homogeneously
through the  tails, but form clumps,  and the densest of them, located
symmetrically  in  the  tails,   are  the  tracers of   the  strongest
gravitational shocks.
\item Mass  loss is highly enhanced with  a ``maximum disk'' model for
the Galaxy. On  the  contrary,  the flattening  of the  dark  halo has
negligible effect on the clusters, for a given rotation curve.
\end{itemize}
Finally,  these  N-body experiments  help  to   understand the  recent
observations  of extended    tidal  tails around   globular   clusters
(Grillmair et al. 1995, Leon et al. 1999): the systematic observations
of the geometry of these  tails should bring  much information on  the
orbit, dynamic, and mass loss history of the clusters.

\acknowledgements{All simulations have  been carried out on  the Crays
C-94 and C-98 of IDRIS, the CNRS computing center at Orsay, France. }

\end{document}